\documentclass[sigconf]{acmart}
\usepackage{soul}
\usepackage{graphicx}
\usepackage{booktabs}
\usepackage{float}
\usepackage{overpic}
\usepackage{algorithm}
\usepackage{algpseudocode}
\usepackage{float}  
\usepackage{subfloat}
\usepackage{stfloats} 
\usepackage[skip=0.5\baselineskip]{caption}
\usepackage{bm}
\usepackage{amsmath}
\usepackage{booktabs}
\usepackage{multirow}
\usepackage{multicol}
\usepackage{enumitem}
\usepackage{subcaption}
\usepackage{threeparttable}
\usepackage{xspace}
\usepackage[normalem]{ulem}
\usepackage{algorithmicx,algorithm}
\usepackage{algpseudocode}
\usepackage{algorithm,algpseudocode}
\usepackage{marvosym}   
\usepackage{colortbl}   
\usepackage{tcolorbox}  
\tcbuselibrary{breakable}   
\usepackage{wasysym}    
\usepackage[figuresright]{rotating} 
\usepackage{pifont} 

\newcommand{\ie}{\emph{i.e.,}\xspace}
\newcommand{\eg}{\emph{e.g.,}\xspace}
\newcommand{\name}{H$^2$Rec}
\AtBeginDocument{%
	\providecommand\BibTeX{{%
			\normalfont B\kern-0.5em{\scshape i\kern-0.25em b}\kern-0.8em\TeX}}}

\copyrightyear{2026}
\acmYear{2026}
\setcopyright{cc}
\setcctype{by-nc-nd}
\acmConference[KDD '26]{Proceedings of the 32nd ACM SIGKDD Conference on Knowledge Discovery and Data Mining V.2}{August 09--13, 2026}{Jeju Island, Republic of Korea}
\acmBooktitle{Proceedings of the 32nd ACM SIGKDD Conference on Knowledge Discovery and Data Mining V.2 (KDD '26), August 09--13, 2026, Jeju Island, Republic of Korea}
\acmDOI{10.1145/3770855.3818490}
\acmISBN{979-8-4007-2259-2/2026/08}
\begin{document}
\title{The Best of Both Worlds: Harmonizing Semantic and Hash IDs for Sequential Recommendation}
\author{Ziwei Liu}
\authornote{Equal Contribution.}
\affiliation{%
  \institution{City University of Hong Kong}
  \city{Hong Kong}
  \country{China}
}
\email{lziwei2-c@my.cityu.edu.hk}

\author{Yejing Wang}
\authornotemark[1]
\affiliation{%
  \institution{City University of Hong Kong}
  \city{Hong Kong}
  \country{China}
}
\email{yejing.wang@my.cityu.edu.hk}

\author{Wanyu Wang}
\affiliation{%
  \institution{City University of Hong Kong}
  \city{Hong Kong}
  \country{China}
}
\email{wanyuwang4-c@my.cityu.edu.hk}

\author{Wang Zejian}
\affiliation{%
  \institution{Tongji University}
  \city{Shanghai}
  \country{China}
}
\email{wangzejian_1215@163.com}

\author{Qidong Liu}
\affiliation{%
  \institution{Xi'an Jiaotong University}
  \city{Xi'an}
  \country{China}
}
\email{liuqidong@xjtu.edu.cn}

\author{Zijian Zhang}
\affiliation{%
  \institution{Jilin University}
  \city{Changchun}
  \country{China}
}
\email{zhangzijian@jlu.edu.cn}

\author{Wei Huang}
\affiliation{%
  \institution{Independent Researcher}
  \city{Beijing}
  \country{China}
}
\email{hwdzyx@gmail.com}

\author{Chong Chen}
\affiliation{%
  \institution{Tsinghua University}
  \city{Beijing}
  \country{China}
}
\email{cstchenc@163.com}

\author{Xiangyu Zhao}\authornote{Corresponding author}
\affiliation{%
  \institution{City University of Hong Kong}
  \city{Hong Kong}
  \country{China}
}
\email{xianzhao@cityu.edu.hk}

\renewcommand{\shortauthors}{Ziwei Liu, et al.}

\begin{abstract}
Conventional Sequential Recommender Systems (SRS) typically assign unique hash IDs (HID) to construct item embeddings, which mainly capture collaborative signals from historical user–item interactions. However, such embeddings are vulnerable in long-tail scenarios where most items are rarely consumed.
Recent methods that incorporate auxiliary information often face noisy collaborative sharing from co-occurrence signals or semantic homogeneity caused by flat dense embeddings. In contrast, Semantic IDs (SID), with their support for code sharing and multi-granular semantic modeling, offer a promising alternative.
Nevertheless, SID-based methods are hindered by a collaborative overwhelming phenomenon: commonly adopted quantization mechanisms compromise the identifier uniqueness needed to model head items, resulting in a performance trade-off between head and tail items.
To address this challenge, we propose \textbf{\name}, a novel framework that harmonizes SID and HID. We design a dual-branch modeling architecture that simultaneously captures the multi-granular semantics of SID while preserving the unique collaborative identity provided by HID. Moreover, we introduce a dual-level alignment strategy to bridge the two representations, enabling effective knowledge transfer and robust preference modeling. 
Extensive offline experiments on three public benchmarks and online experiments on a large-scale commercial platform demonstrate that \name~ achieves a better balance between head and tail recommendation quality and consistently outperforms existing baselines. The implementation code is publicly available\footnote{\url{https://github.com/Applied-Machine-Learning-Lab/KDD26_H2Rec}}.
\end{abstract}



\maketitle
\section{Introduction}\label{sec:intro}
\begin{figure}[!t]
    \centering
    \includegraphics[width = \linewidth]{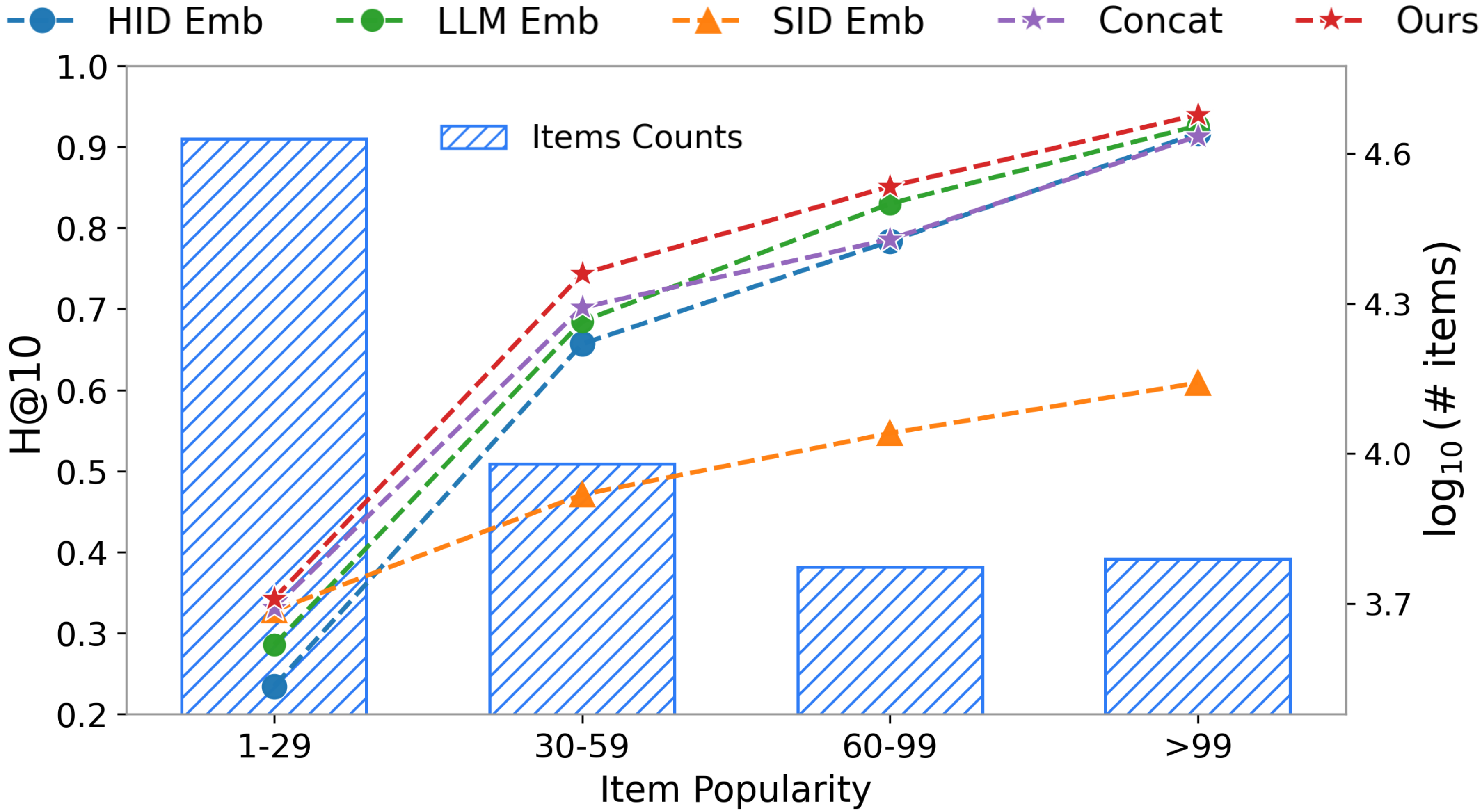}
    \caption{Performance comparison with SASRec on Yelp (item popularity breakdown).The marker indicates the method type (\ding{108}: HID-based, \ding{115}: SID-based, \ding{72}: hybrid).}
    \label{fig:prel_exp}
\end{figure}

Sequential recommender systems (SRS) aim to predict users' next interactions based on their historical behaviors and are widely deployed across digital platforms such as e-commerce~\cite{e-commerce,zhao2020jointly,wang2022autofield,zhang2024m3oe} and streaming media~\cite{stream}. In conventional SRS frameworks, \eg BERT4Rec~\cite{Bert4rec}, GRU4Rec~\cite{GRU4Rec}, and SASRec~\cite{SASRec}, items are assigned unique identifiers (IDs) to enable hash-based lookup~\cite{sparsity}. These hash IDs (HID) are then mapped to high-dimensional embeddings that encode collaborative information from historical interactions~\cite{zhao2023embedding}, allowing the SRS to effectively learn users’ preferences and provide accurate Top-K recommendations.

However, HID-based SRS suffer from a long-standing drawback: the long-tail item problem~\cite{long-tail_item}. As illustrated in Figure~\ref{fig:prel_exp}, HID embeddings (the blue line) degrade significantly for items with sparse interactions (denoted as tail items), which account for more than 70\% of the whole item set. This observation confirms that traditional collaborative filtering methods~\cite{SR_survey} struggle to learn reliable representations for tail items due to insufficient data~\cite{sparsity}. 
To address this issue, prior studies~\cite{Melt,CITIES} exploit co-occurrence patterns and enhance tail items by leveraging the collaborative information from popular items that the same user has interacted with. Although effective to some extent, these approaches overlook the semantic information embedded in item descriptions. Moreover, co-occurrence signals can be unreliable in real-world~\cite{longtail_session}. For example, accidental clicks produce misleading collaborative information, causing tail items to inherit noise from semantically unrelated popular items. This leads to the \ding{182} \textbf{Noisy Collaborative Sharing}.

Recent advances in large language models (LLM) provide a promising avenue to mitigate this issue by enriching HID embeddings with semantic information~\cite{llmsurvey,surveyllmesr}. Current LLM-based item encoders extract semantic features from items' textual attributes, and leverage LLM's pretrained world knowledge to further enhance the representations~\cite{LLM-ESR,LLMemb,llmembeddinglearning_SR,gao2025llm4rerank,zhang2025notellm}. We take the powerful LLM-ESR~\cite{LLM-ESR} as a baseline in our preliminary experiment, denoted as ``LLM Emb''. As shown in the green line of Figure~\ref{fig:prel_exp}, it achieves an improved performance across all item-popularity groups. However, in this paradigm, LLM compresses all textual information into a single dense vector. We argue that this ``flat'' representation induces a single-granularity bottleneck. Here, we define the semantic granularity as the level of abstraction at which item semantics are expressed. A single dense embedding inevitably entangles coarse-grained semantics with fine-grained nuances, making it difficult to distinguish subtle differences among semantically similar items. This limitation gives rise to the \ding{183} \textbf{Semantic Homogeneity}. 

To simultaneously address these problems, recent work~\cite{GR_survey1,GR_survey2,liu2026conditional} has turned to Semantic IDs (SID) for recommendation. Unlike HID, SID are generated by decomposing dense semantic embeddings (\eg from LLMs) into discrete code sequences through vector quantization techniques such as RQ-VAE~\cite{RQVAE}. Crucially, the quantization hierarchy unfolds the dense embedding into multi-granular semantic views. Each residual level corresponds to a distinct abstraction of semantics, forming multiple implicit views rather than explicit metadata such as category or brand. This structure is theoretically appealing: shared codes at the same layer aggregate collaborative signals among semantically related items, mitigating \textbf{Noisy Collaborative Sharing}. Meanwhile, multi-granular codes offer finer semantic distinctions, alleviating the \textbf{Semantic Homogeneity}.

Despite these advantages, existing SID-based methods~\cite{ADA-SID,MME-SID,CCFRec,VQRec,PCRCA,SMILE,Meta_concat} typically replace HID with SID or fuse them using simple concatenation or contrastive learning. Such strategies fail to fully exploit the potential of SID and are hindered by the \textbf{Collaborative Overwhelming} phenomenon~\cite{GR_survey1,GR_survey3}. 
This phenomenon stems from the quantization process, which inevitably introduces code collisions (\eg multiple items share the same SID)~\cite{ID_collision}. These collisions compromise the uniqueness of ID-item mappings, confusing the model with inflated user-item connections. This effect is particularly pronounced for `head' items with abundant interactions, as evidenced by the orange line in Figure~\ref{fig:prel_exp}. Consequently, Figure~\ref{fig:prel_exp} reveals an intrinsic trade-off: pure SID lack the uniqueness required for head items, while pure HID lack the semantic information to benefit tail items. This underscores the need for a harmonized framework that integrates the strengths of both paradigms to achieve the best of two worlds.

In this paper, we propose a framework with \textbf{\uline{H}}armonized Semantic and \textbf{\uline{H}}ash IDs for Sequential \textbf{\uline{Rec}}ommendation (\textbf{\name}). Our framework employs a dual-branch architecture with dual-level alignment. Specifically, the SID branch includes a multi-granularity fusion network that constructs fine-grained semantic representations, while the HID branch incorporates a multi-granularity cross-attention network that selectively injects semantic signals into HID embeddings. To further harmonize the HID and SID representations, we introduce a Dual-level Alignment Strategy: a code-guided alignment loss at the item level, and a masked sequence granularity loss at the user level. It is important to note that our framework is quantization-agnostic and model-agnostic, making it compatible with different quantization mechanisms and SRS architectures.
Our contributions are summarized as follows:
\begin{itemize}[leftmargin=*]
    \item We identify the \textbf{Collaborative Overwhelming} phenomenon in SID-based methods, revealing a fundamental seesaw between the identifier uniqueness required for head items and the semantic needed for tail items.
    \item We propose \textbf{\name}, a dual-branch framework that harmonizes HID and SID. By combining our multi-granularity modules with the Dual-level Alignment Strategy, \name~ effectively addresses the limitations of each identifier type.
    \item Extensive experiments online and offline demonstrate the effectiveness of \name~ , which achieves robust performance no matter for the head or tail items.
\end{itemize}
\section{Preliminary}
\subsection{Problem Definition}
The sequential recommendation aims to predict the next interacted items based on the historical record. Specifically, we can derive the interaction sequence for a specific user $u \in \mathcal{U}$, denoted as $\mathcal{S}_u=\{v_1,\cdots,v_{N}\}$. where $v_i \in \mathcal{V}$ represents the interacted item i in the item set $\mathcal{V}$, and $N$ represents the length of the interaction sequence. Correspondingly, we define the problem of SRS as:
\begin{equation}
    \label{equ:problem_SRS}
    \arg\max_{v_i \in \mathcal{V}} P(v_{N+1}= v_i |\mathcal{S}_u)
\end{equation}
\subsection{Semantic IDs Generation}
\label{sec:semantic IDs}
Vector Quantization has been widely used to generate the SID from the original embedding~\cite{GR_survey3}. In our settings, we directly take the widely used Residual Quantized Variational Autoencoder (RQ-VAE)~\cite{RQVAE} as our SID generator, following previous work~\cite{Tiger}. Specifically, for a semantic embedding $\boldsymbol{e}_{LLM}$ derived from a language model encoder, we leverage RQ-VAE to quantize it into $L$ semantic codes, where $L$ denotes the number of code layers. Consequently, the item $v_i$ can be represented as a tuple $C_i=(c^1_i, \dots, c^L_i)$, where $c_i^l$ is a single code whose embedding $\boldsymbol{e}_{c_i}^l$ corresponds to the slice of the codebook embedding $\boldsymbol{E}_{C}^l$.
\section{Methodology}
In this section, we introduce the architecture of the proposed \name. We begin with an overview of the framework, followed by detailed illustrations of its dual-branch structure and dual-level alignment. At last, the training and inference procedure is specified.
\begin{figure*}[!t]
	\centering
	\includegraphics[width = \linewidth]{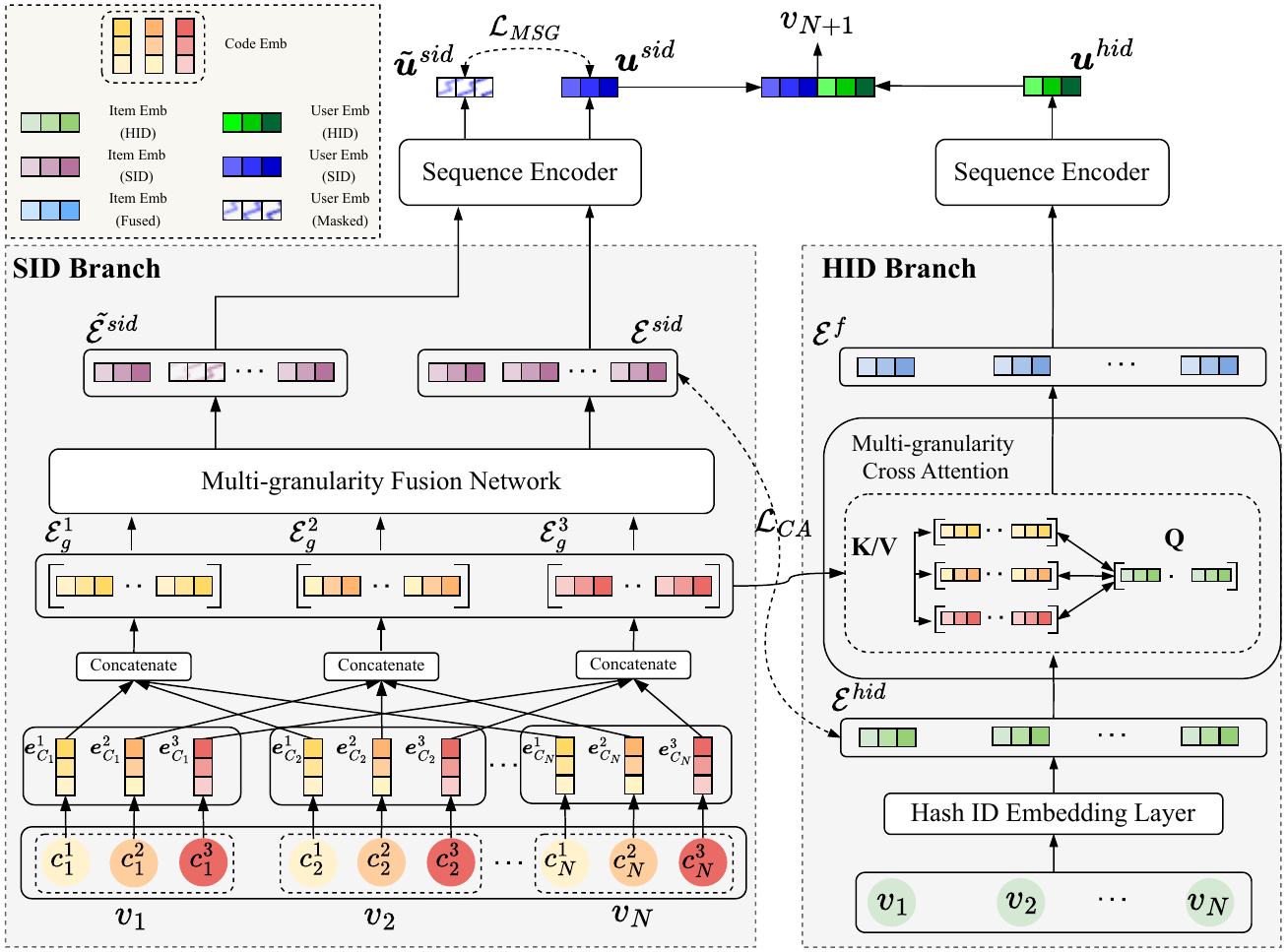}
	\caption{The framework overview for \name. In this figure, we set $L=3$ for better illustration.}
        \label{fig:Overall}
\end{figure*}

\subsection{Overview}
The overview of our proposed framework is illustrated in Figure~\ref{fig:Overall}. 
To harmonize the two types of identifiers, we first devise a \textbf{Dual-branch Modeling} architecture. In the \textbf{SID Branch}, we employ an RQ-VAE framework, which maps textual attributes into $L$ discrete code levels, to construct our Semantic code embeddings. Then, to construct multi-granular representations, the generated code embeddings of the item sequence are concatenated based on their corresponding code levels, forming $L$-granularity sequences. 
With the proposed \textbf{Multi-granularity Fusion Network}, these sequences are then adaptively aggregated into a final fine-grained SID embedding sequence. Finally, a sequence encoder is employed to generate the user representation for SID branch, denoted as $\boldsymbol{u}^{sid}$. In the \textbf{HID branch}, we first leverage a learnable embedding layer to derive the HID item embedding. After that, we introduce the \textbf{Multi-granularity Cross Attention Network}. Using the unique HID as the query, this module selectively incorporates essential semantic information from SID while preserving the uniqueness of head items, successfully generating an information-fused embedding sequence. Finally, another sequence encoder is applied on the fused embedding sequence to derive the user representation $\boldsymbol{u}^{hid}$ in the HID branch.

To further optimize the whole architecture, a \textbf{Dual-level Alignment Strategy} is proposed.
At the item level, we design a \textbf{Code-Guided Alignment} loss that aligns the collaborative and semantic spaces with well-designed sample selection. This module allows tail items to borrow high-quality collaborative signals from semantically similar head items without introducing noise. At the user level, to ensure robust user representation, we randomly mask a subset of granularities to obtain the masked user representation $\tilde{\boldsymbol{u}}^{sid}$. Then, we construct a \textbf{Masked Sequence Granularity Loss} to enhance internal correlations among multi-granular semantics.

At last, the mutually enhanced HID and SID user representations, \ie $\boldsymbol{u}^{hid}$ and $\boldsymbol{u}^{sid}$, are fused to derive the recommendation scores.

\subsection{Dual-branch Modeling}
In this section, to mitigate the \textbf{Collaborative Overwhelming} phenomenon~\cite{GR_survey1,GR_survey3}, and leverage the complementary benefits of the two different IDs, we present our dual-branch modeling, featuring the SID branch and the HID branch.

\subsubsection{\textbf{SID Branch}}\label{sec:sid}
To acquire the semantic code embeddings for each item, we first derive semantic embeddings from the item’s textual attributes by leveraging Large Language Models (LLM)~\cite{surveyllmesr}. Specifically, we convert item attributes into textual instructions and obtain the representations of all items, denoted as $\boldsymbol{E}_{LLM} \in \mathbb{R}^{|\mathcal{V}| \times d_{llm}}$, using open-resource LLM~\cite{llama,qwen} or a public API such as text-embedding-ada-002\footnote{\url{https://platform.openai.com/docs/guides/embeddings}}. Subsequently, the RQ-VAE framework is trained to derive the semantic codes $C_i$ together with their corresponding codebook embedding matrices $\boldsymbol{E}_C$, as described in Section~\ref{sec:semantic IDs}. The semantic code embeddings for item $i$ are then retrieved from the $L$ codebook embedding matrices through look-up operations, forming the sequence $[\boldsymbol{e}_{C_i}^1,\dots,\boldsymbol{e}_{C_i}^L]$.

Following this, we introduce the multi-granularity fusion network in our SID branch to generate fine-grained SID representations. In this module, we first organize the code embeddings of the user's interaction sequence by their code levels. The embedding sequence for a specific code level $l$ is defined as $\mathcal{E}_{g}^l = [\boldsymbol{e}_{C_1}^{l},\boldsymbol{e}_{C_2}^{l},\dots,\boldsymbol{e}_{C_N}^{l}]$, where $N$ represents the sequence length.

Since user intent shifts dynamically, a fixed integration strategy is suboptimal. Therefore, we design an adaptive mechanism that leverages the user's interaction context to assign importance weights to different code-level granularities. Formally, we use the embedding of the user's last interacted item from the HID embedding sequence, denoted as $\boldsymbol{e}^{hid}_N \in \mathbb{R}^d$, to serve as the query anchor for the current intent. To ensure training stability and incorporate the prior knowledge that coarse-grained semantics are generally more robust, we introduce a learnable residual bias vector $\boldsymbol{b}_{prior} \in \mathbb{R}^L$. The unnormalized importance scores $\boldsymbol{s} \in \mathbb{R}^L$ are computed via a Multi-Layer Perceptron (MLP) as follows:
\begin{equation}
\label{equ:score}
\boldsymbol{s} = \boldsymbol{W}_2 \left( \sigma \left( \boldsymbol{W}_1 [\boldsymbol{e}^{hid}_N ; \boldsymbol{b}_{prior}] + \boldsymbol{b}_1 \right) \right) + \boldsymbol{b}_2 + \boldsymbol{b}_{prior}
\end{equation}
where $[\cdot ; \cdot]$ represents the concatenation operation, and $\boldsymbol{W}_1 \in \mathbb{R}^{d_{h} \times (d+L)}$, $\boldsymbol{W}_2 \in \mathbb{R}^{L \times d_{h}}$, $\boldsymbol{b}_1 \in \mathbb{R}^{d_h}$, and $\boldsymbol{b}_2 \in \mathbb{R}^L$ are learnable parameters. The residual addition of $\boldsymbol{b}_{prior}$ explicitly encourages the model to respect the hierarchical structure of SID generated by RQ-VAE. Finally, we normalize the scores into importance weights $\boldsymbol{\alpha}$ and derive the final SID item embedding sequence $\mathcal{E}^{sid} = [\boldsymbol{e}^{sid}_1,\dots,\boldsymbol{e}^{sid}_N]$ by aggregating the granularities:
\begin{equation}
\label{equ:weights}
    \alpha_l = \frac{\exp(s_l)}{\sum_{k=1}^L \exp(s_k)}, \quad \mathcal{E}^{sid} = \sum_{l=1}^{L} \alpha_l \cdot \mathcal{E}_{g}^l
\end{equation}
We then employ an independent sequence encoder $f_{\theta^{sid}}$ to acquire the final user representation $\boldsymbol{u}^{sid}$ for the SID branch.

\subsubsection{\textbf{HID Branch}}\label{sec:hid}
First, we derive the item embedding for the HID branch by employing a learnable item embedding layer and updating it by absorbing the collaborative information from historical user-item interactions. Let $\boldsymbol{E}_{hid} \in \mathbb{R}^{|\mathcal{V}|\times d}$ represent the HID embedding layer. Then, the item embeddings sequence $\mathcal{E}^{hid} = [\boldsymbol{e}^{hid}_1,\dots,\boldsymbol{e}^{hid}_{N}]$ can be derived by extracting the corresponding rows from the embedding layer $\boldsymbol{E}_{hid}$.

To inject nuanced, multi-granular semantic information back into the HID item embedding, we propose the Multi-Granularity Cross Attention mechanism. Formally, we treat the HID item embedding sequence $\mathcal{E}^{hid}$ as the query anchor, while treating the multi-granular SID item embedding sequences $\mathcal{E}^l_g$ (defined in Section~\ref{sec:sid}) as the key-value pairs. This strategy allows the model to utilize the unique HID as a query to selectively retrieve and inject semantic information from the SID.

Specifically, for each code level $l$, we first project the collaborative and semantic sequences into distinct subspaces using learnable weight matrices $\boldsymbol{W}^Q, \boldsymbol{W}^K, \boldsymbol{W}^V \in \mathbb{R}^{d \times d}$, formulated as:
\begin{equation}
\boldsymbol{Q} = \mathcal{E}^{hid}\boldsymbol{W}^Q, \quad \boldsymbol{K}_l = \mathcal{E}_{g}^l\boldsymbol{W}^K, \quad \boldsymbol{V}_l = \mathcal{E}_{g}^l\boldsymbol{W}^V
\end{equation}
Based on these projections, we derive the final fused embedding sequence $\mathcal{E}^{f}$ by aggregating the granularity-specific attention outputs, weighted by the user's intent-aware scores $\alpha_l$ derived in Equation~\eqref{equ:weights}. Crucially, a residual connection is employed to add back the original HID item embedding sequence:
\begin{equation}
\label{equ:MCA}
\mathcal{E}^{f} = \sum_{l=1}^{L} \alpha_l \cdot \left( \operatorname{softmax}\left(\frac{\boldsymbol{Q}\boldsymbol{K}_l^\top}{\sqrt{d}}\right)\boldsymbol{V}_l \right) + \mathcal{E}^{hid}
\end{equation}
Finally, a sequence encoder $f_{\theta^{hid}}$ is employed on the fused item embedding sequence to derive the comprehensive user representation $\boldsymbol{u}^{hid}$ for the HID branch.

\subsection{Dual-level Alignment}
To further enhance the representation ability of each branch, we propose a dual-level alignment approach, featuring a code-guided alignment loss at the item level and a masked sequence granularity loss at the user level, further alleviating the \textbf{Noisy Collaborative Sharing} and \textbf{Semantic Homogeneity}, respectively.

\subsubsection{\textbf{Code-guided Alignment Loss}}\label{sec:CA}
To mitigate {Noisy Collaborative Sharing}~\cite{longtail_session}, we aim to align the semantic and collaborative spaces. A straightforward solution is to adopt a contrastive learning objective~\cite{contrastivel} that pulls the SID item embedding $\boldsymbol{e}^{sid}_i$ and HID item embedding $\boldsymbol{e}^{hid}_i$ of the same item $i$ closer while pushing them away from negative samples. This 1-to-1 alignment explicitly transfers the collaborative signals unique to item $i$ into its semantic representation, formulated as:
\begin{equation}
\mathcal{L}_{align} = -\frac{1}{B} \sum^{B}_{i=1} \log \frac{\exp(\cos(\boldsymbol{e}^{sid}_i,\boldsymbol{e}^{hid}_i)/\tau)}{\sum_{j=1}^{B} \mathbb{I}_{[j \neq i]} \exp(\cos({\boldsymbol{e}^{sid}_i},{\boldsymbol{e}^{hid}_j})/\tau)}
\end{equation}
where $B$ denotes the batch size, $\cos(\cdot,\cdot)$ represents cosine similarity, and $\tau$ is the temperature coefficient. The indicator function $\mathbb{I}_{[j \neq i]}$ explicitly excludes the item itself from the denominator, ensuring the contrast is computed strictly against other items in the batch.

However, this strict 1-to-1 mapping is insufficient for our goal: enabling long-tail items to implicitly ``borrow'' high-quality collaborative signals from semantically similar head items. To address this, we expand the objective to a 1-to-many code-guided alignment. We construct a unified positive set $\mathcal{P}(i)$ for item $i$ by incorporating: (1) $\mathcal{P}_C(i)$, items sharing $p$ levels of semantic codes with item $i$; and (2) $\mathcal{P}_H(i)$, items appearing within the co-occurrence context window $o$. Let $\mathcal{P}(i) = \{i\} \cup \mathcal{P}_C(i) \cup \mathcal{P}_H(i)$. We then reconstruct the objective to maximize the cumulative similarity between the anchor and all the pre-defined positive samples:
\begin{equation}
\label{equ:lca}
\mathcal{L}_{CA}^1 = - \frac{1}{B} \sum_{i=1}^{B} \log \frac{\sum_{k \in \mathcal{P}(i)} \exp(\cos(\boldsymbol{e}^{sid}_i, \boldsymbol{e}^{hid}_k)/\tau)}{\sum_{j=1}^{B} \mathbb{I}_{[j \neq i]} \exp(\cos(\boldsymbol{e}^{sid}_i, \boldsymbol{e}^{hid}_j)/\tau)}
\end{equation}
To obtain the second part of the alignment loss, we exchange the roles of $\boldsymbol{e}^{sid}$ and $\boldsymbol{e}^{hid}$, yielding $\mathcal{L}_{CA}^2$. The final alignment loss is: $\mathcal{L}_{CA} = \mathcal{L}_{CA}^1 +\mathcal{L}_{CA}^2$.

\subsubsection{\textbf{Masked Sequence Granularity Loss}}
To further address the Semantic Homogeneity~\cite{llmembeddinglearning_SR} and thereby enhance the consistency, we proposed the masked sequence granularity loss.

For each user interaction sequence, we construct two views: a global view and a granularity-masked view. In the masked view, we randomly sample a granularity index $m \in [1, L]$ and replace the embedding sequence at this level with a learnable mask token, resulting in $\tilde{\mathcal{E}}^{sid}$. Both views are processed to derive the global representation $\boldsymbol{u}^{sid}$ and the masked representation $\tilde{\boldsymbol{u}}^{sid}$.

To ensure the model can implicitly infer missing semantic information, we maximize the mutual information between these representations. Formally, for a batch of $N$ users, we minimize the distance between the positive pair $(\boldsymbol{u}^{sid}_i, \tilde{\boldsymbol{u}}^{sid}_i)$ while pushing apart negative pairs from other users in the batch:
\begin{equation}
\label{equ:lmsg}
\mathcal{L}_{MSG}^1 = - \frac{1}{N} \sum_{i=1}^{N} \log \frac{\exp(\operatorname{cos}(\boldsymbol{u}^{sid}_i, \tilde{\boldsymbol{u}}^{sid}_i) / \tau)}{\sum_{k=1}^{N} \exp(\operatorname{cos}(\boldsymbol{u}^{sid}_i, \tilde{\boldsymbol{u}}^{sid}_k) / \tau)}
\end{equation}
where the denominator index $k$ iterates over all $N$ users. This ensures that for any $k \neq i$, $\tilde{\boldsymbol{u}}^{sid}_k$ serves as a negative sample, preventing the objective from collapsing into a trivial solution. Exchanging the positions of $\tilde{\boldsymbol{u}}^{sid}$ and ${\boldsymbol{u}}^{sid}$ yields $\mathcal{L}_{MSG}^2$. The final loss is: $\mathcal{L}_{MSG}=\mathcal{L}_{MSG}^1 + \mathcal{L}_{MSG}^2 $.

\subsection{Training and Inference}
\subsubsection{\textbf{Training}} 
As described in Section~\ref{sec:hid} and Section~\ref{sec:sid}, we first obtain the embedding sequences $\mathcal{E}^{hid}$ and $\mathcal{E}^{sid}$ for each branch and leverage two individual sequence encoder to derive the final $\boldsymbol{u}^{sid}$ and $\boldsymbol{u}^{hid}$.  After that, we can compute the probability of recommending item $j$ to user $u$ based on the fused representations:
\begin{equation}
\label{equ:rec}
    P(v_{N+1} = v_j | \mathcal{S}_u) = [\boldsymbol{e}_j^{sid}:\boldsymbol{e}_j^{hid}]^\top [\boldsymbol{u}^{sid}:\boldsymbol{u}^{hid}]
\end{equation}
where $:$ denotes the concatenation operator. Based on these scores, we optimize the model using a pairwise ranking loss~\cite{Cross-entropy}:
\begin{equation}
\label{equ:recloss}
    \mathcal{L}_{rec}=-\sum_{u\in\mathcal{U}}\sum_{j=1}^{N}\operatorname{log}\sigma(P(v_{j+1}=v^+|\mathcal{S}_u)-P(v_{j+1}=v^-|\mathcal{S}_u))
\end{equation}
where $v^+$ and $v^-$ respectively represent the ground-truth item and the corresponding negative item.

Finally, the overall training objective combines the main recommendation loss with the code-guided alignment Loss and the masked sequence granularity loss:
\begin{equation}
    \mathcal{L}=\mathcal{L}_{rec} + \beta \cdot \mathcal{L}_{CA} + \gamma \cdot \mathcal{L}_{MSG}
\end{equation}
where $\beta$ and $\gamma$ are hyperparameters controlling the contributions of the two auxiliary objectives.
\subsubsection{\textbf{Inference}}
During inference, since the SID and the associated codebook embeddings are cached in advance, we directly compute probability using Equation~\eqref{equ:rec} to derive the recommended items. For detailed procedures, please refer to \textbf{Appendix}~\ref{sec:train_infer}. 

\section{Experiment}
\label{sec:experiment}
\subsection{Experiment Settings}
\subsubsection{\textbf{Datasets}}
We apply three real-world datasets to evaluate the effectiveness of our proposed \name, \ie Yelp, Amazon Beauty, and Amazon Instrument. The details can be found in \textbf{Appendix}~\ref{sec:dataset}.
\subsubsection{\textbf{Baselines}}
To validate the superiority of \name, we conduct our comparison experiment with various up-to-date baselines under three categories.
\begin{itemize}[leftmargin=*]
    \item \textbf{HID Embedding}: BERT4Rec~\cite{Bert4rec}, SASRec~\cite{SASRec}, MELT~\cite{Melt}, and LLM-ESR~\cite{LLM-ESR}. 
    \item \textbf{SID Embedding}: PG-SID~\cite{prefixgram_SID}, SPM-SID~\cite{SPM_SID}, CCFRec~\cite{CCFRec} and PSRQ+MCCA~\cite{PSRQ}, which is denoted as PSRQ for simplicity.
    \item \textbf{Hybrid Embedding}: URL4DR~\cite{Meta_concat}, MME-SID~\cite{MME-SID}, PCR-CA~\cite{PCRCA}, and SMILE~\cite{SMILE}.
\end{itemize}
\subsubsection{\textbf{Implementation Details}}
We implement all our experiments on a single RTX 4090 24G GPU, while the basic software requirements are Python 3.11 and PyTorch 2.4-2204. $\beta$ and $\gamma$ are all searched from $\{0.1,0.3,0.5,0.7,0.9\}$. Our code is available online at the repository \footnote{\url{https://github.com/Applied-Machine-Learning-Lab/KDD26_H2Rec}}. For more details, please refer to \textbf{Appendix}~\ref{sec:imple}. 
\subsubsection{\textbf{Evaluation Metrics}}
\label{sec:metric}
To provide a comprehensive view of our framework, following existing research~\cite{LLM-ESR,long-tail_item,longtail_session}, we split the items into tail and head groups by selecting the top $20\%$ items as head items and regarding the rest as tail items. For detailed evaluation for each group, we adopt the commonly used \textit{Hit Rate} and \textit{Normalized Discounted Cumulative Gain}, all truncated at 10, denoted as ${H@10}$ and ${N@10}$ respectively. To ensure the robustness of the experimental results, all the presented results are the average of the three runs with random seeds $\{42, 43, 44\}$.
\begin{table*}[t]
\centering
\caption{Overall performance of the proposed \name. The best results are bold, and the second-best are underlined. ``*'' indicates the improvements are statistically significant (i.e., two-sided t-test with \(p<0.05\) ) over baselines.}
\label{tab:overall_performance}
\resizebox{\textwidth}{!}{%
\begin{tabular}{ccl|cccc|cccc|cccc|c|c}
\toprule
\multirow{2}{*}{\textbf{Dataset}} & \multirow{2}{*}{\textbf{Group}} & \multirow{2}{*}{\textbf{Metric}} & 
\multicolumn{4}{c|}{\textbf{HID Emb}} & \multicolumn{4}{c|}{\textbf{SID Emb}} & \multicolumn{4}{c|}{\textbf{Hybrid Emb}} & \textbf{Ours} & \textbf{Improv.} \\
\cmidrule(lr){4-7} \cmidrule(lr){8-11} \cmidrule(lr){12-15} \cmidrule(lr){16-16} \cmidrule(lr){17-17}
 & & & BERT4Rec & SASRec & MELT & LLM-ESR & SFM-SID & PG-SID & CCFRec & PSRQ & URL4DR & MME-SID & PCR-CA & SMILE & \textbf{\name} & \% \\
\midrule

\multirow{6}{*}{Yelp} 
 & \multirow{2}{*}{Overall} 
   & H@10   & 0.5314 & 0.5940 & 0.6101 & \underline{0.6573} & 0.4727 & 0.4881 & 0.5947 & 0.5438 & 0.6402 & 0.6431 & 0.6447 & 0.6487 & \textbf{0.6692}$^*$ & 1.81\% \\
 & & N@10 & 0.3147 & 0.3601 & 0.3394 & \underline{0.4102} & 0.3148 & 0.3251 & 0.3694 & 0.3422 & 0.3776 & 0.3884 & 0.3971 & 0.3988 & \textbf{0.4272}$^*$ & 4.14\% \\
 \cmidrule{2-17}
 & \multirow{2}{*}{Tail} 
   & H@10   & 0.0177 & 0.1175 & 0.1223 & 0.1802 & 0.2441 & 0.2492 & 0.2478 &\underline{0.2543} &  0.1957 & 0.2215 & 0.2032 & 0.2155 & \textbf{0.2693}$^*$ & 5.90\% \\
 & & N@10 & 0.0068 & 0.0588 & 0.0599 & 0.0676 & 0.1162 & 0.1186 & 0.1171 & \underline{0.1210}  & 0.0914 & 0.1103 & 0.0954 & 0.1057 & \textbf{0.1306}$^*$ & 7.93\% \\
 \cmidrule{2-17}
 & \multirow{2}{*}{Head} 
   & H@10   & 0.6919 & 0.7413 & 0.7790 & \underline{0.8059} & 0.5218 & 0.5334 & 0.7071 & 0.6398 & 0.7699 & 0.7702 & 0.7748 & 0.7820 & \textbf{0.8324}$^*$ & 3.29\% \\
 & & N@10 & 0.3876 & 0.4592 & 0.4745 & \underline{0.5122} & 0.3079 & 0.3147 & 0.4243 & 0.3721 & 0.4735 & 0.4823 & 0.4860 & 0.4916 & \textbf{0.5483}$^*$ & 7.05\% \\
\midrule

\multirow{6}{*}{Beauty} 
 & \multirow{2}{*}{Overall} 
   & H@10   & 0.3992 & 0.4401 & 0.4890 & \underline{0.5544} & 0.3715 & 0.3836 & 0.4398 & 0.4038 & 0.5464 & 0.5509 & 0.5539 & 0.5543 & \textbf{0.5742}$^*$ & 3.57\% \\
 & & N@10 & 0.2401 & 0.3043 & 0.3357 & \underline{0.3702} & 0.2617 & 0.2703 & 0.3021 & 0.2845 & 0.3675 & 0.3617 & 0.3673 & 0.3698 & \textbf{0.3957}$^*$ & 6.89\% \\
 \cmidrule{2-17}
 & \multirow{2}{*}{Tail} 
   & H@10   & 0.0123 & 0.0921 &  0.1536 & 0.2198 & 0.2208 & 0.2254 & 0.2238 & \underline{0.2300}  & 0.1967 & 0.2177 & 0.2048 & 0.2103 & \textbf{0.2557}$^*$ & 11.17\% \\
 & & N@10 & 0.0052 & 0.0675 & 0.0877 & 0.1074 & 0.1355 & 0.1383 & 0.1323 & \underline{0.1411}  & 0.1105 & 0.1404 & 0.1342 & 0.1357 & \textbf{0.1524}$^*$ & 8.01\% \\
 \cmidrule{2-17}
 & \multirow{2}{*}{Head} 
   & H@10   & 0.4988 & 0.5291 & 0.5815 & 0.6377 & 0.4388 & 0.4485 & 0.5099 & 0.4875 & 0.6299 & 0.6393 & 0.6344 & \underline{0.6419} & \textbf{0.6502}$^*$ & 1.29\% \\
 & & N@10 & 0.2971 & 0.4007 & 0.4106 & 0.4289 & 0.3251 & 0.3323 & 0.3887 & 0.3612 & 0.4300 & 0.4379 & 0.4277 & \underline{0.4401} & \textbf{0.4538}$^*$ & 3.11\% \\
\midrule

\multirow{6}{*}{Instrument} 
 & \multirow{2}{*}{Overall} 
   & H@10   & 0.4601 & 0.5057 & 0.5510 & 0.5881 & 0.4312 & 0.4453 & 0.5078 & 0.4687 & 0.6005 & 0.6044 & \underline{0.6072} & 0.6070 & \textbf{0.6184}$^*$ & 1.84\% \\
 & & N@10 & 0.3213 & 0.3442 & 0.3622 & 0.3809 & 0.2908 & 0.3003 & 0.3376 & 0.3161 & 0.4024 & 0.4027 & 0.4034 & \underline{0.4056} & \textbf{0.4153}$^*$ & 2.39\% \\
 \cmidrule{2-17}
 & \multirow{2}{*}{Tail} 
   & H@10   & 0.0199 & 0.0489 & 0.0766 & 0.0998 & 0.2058 & 0.2101 & 0.2099 & \underline{0.2144}  & 0.1605 & 0.2044 & 0.1827 & 0.2001 & \textbf{0.2382}$^*$ & 11.10\% \\
 & & N@10 & 0.0143 & 0.0257 & 0.0459 & 0.0549 & 0.0990 & 0.1010 & 0.0907 & \underline{0.1031}  & 0.0828 & 0.0991 & 0.1025  & 0.1030 & \textbf{0.1233}$^*$ & 11.88\% \\
 \cmidrule{2-17}
 & \multirow{2}{*}{Head} 
   & H@10   & 0.5028 & 0.5806 & 0.6188 & 0.6676 & 0.5569 & 0.5693 & 0.5629 & 0.6188 & 0.6643 & 0.6646 & 0.6658  & \underline{0.6701} & \textbf{0.6832}$^*$ & 1.95\% \\
 & & N@10 & 0.3190 & 0.3764 & 0.4237 & 0.4522 & 0.3813 & 0.3898 & 0.4192 & 0.4237 & 0.4483 & 0.4498 & 0.4502 & \underline{0.4543} & \textbf{0.4638}$^*$ & 2.09\% \\
\bottomrule
\end{tabular}%
}
\end{table*}

\subsection{Offline Performance}
In this section, we evaluate the performance of our model across different item popularity groups on three public datasets. As shown in \textbf{Table}~\ref{tab:overall_performance}, \name~achieves substantial improvements over state-of-the-art baselines, with relative gains ranging from $1.29\%$ to $11.88\%$ across multiple evaluation metrics.

\subsubsection{\textbf{Overall}}
From \textbf{Table}~\ref{tab:overall_performance}, following \name, the Hybrid Embedding methods (\ie SMILE, PCR-CA, and MME-SID) and the LLM-based method (\ie LLM-ESR) generally yield the second-best performance. This observation suggests that although integrating LLM-derived knowledge into HID or simply combining multiple embeddings can improve performance, these strategies remain insufficient without explicit, fine-grained alignment between the collaborative and semantic spaces.

For SID Embedding baselines, we observe a clear performance disparity. Although the recently proposed SFM-SID and PG-SID adopt more sophisticated mechanisms to enhance semantic representations, they generally underperform compared to CCFRec and PSRQ. This indicates that strengthening semantic quality alone is insufficient without a design oriented toward recommendations. 

Among SID-based methods, CCFRec performs notably comparable to the HID-based SASRec. Its advantage lies in progressively incorporating semantic embeddings across multiple attribute dimensions, yielding finer item distinctions than the sub-vector grouping strategies in SFM-SID and PG-SID. Nevertheless, despite these improvements, single-view methods (first seven columns) are consistently inferior to hybrid approaches (last four columns), further demonstrating the necessity of integrating SID and HID representations in a unified framework.

\subsubsection{\textbf{Popularity Breakdown}}
To deeply investigate how our \name~ benefits when dealing with the long-tail problem, we conduct a fine-grained popularity breakdown analysis by segregating the evaluation into \textit{Head} and \textit{Tail} item groups. 
4
Specifically, HID Embedding-based methods (\eg SASRec and LLM-ESR) excel on Head items due to abundant interaction data, yet suffer drastic performance declines on the Tail group, revealing their reliance on dense collaborative signals.
Conversely, SID-based methods (\eg PSRQ and CCFRec) effectively address sparsity for Tail items by leveraging semantic information for collaborative sharing. For example, PSRQ and CCFRec significantly surpass SASRec on Tail items. However, their performance on the Head group remains limited, primarily due to the inherently coarse-grained nature of SID and the ID collisions.

In contrast, \name~ eliminates this trade-off. It achieves the best results in the Tail group, substantially outperforming the SID-based baselines, while simultaneously maintaining state-of-the-art performance on the Head group compared to the HID-based methods. This demonstrates that \name~ successfully transfers semantic knowledge to cold items without compromising the model’s capability to capture fine-grained collaborative information for popular items, illustrating the effectiveness of our proposed modules.
\begin{table}[t]
  \centering
  \caption{Performance comparison on the Yelp Dataset. ``FN'' denotes the fusion network, and ``MCA'' is the multi-granularity cross attention. Best results are in bold.}
  \label{tab:ablation}
  \resizebox{\linewidth}{!}{
  \begin{tabular}{lcccccc}
    \toprule
    \multirow{2}{*}{\textbf{Variants}} & \multicolumn{2}{c}{\textbf{Overall}} & \multicolumn{2}{c}{\textbf{Tail}} & \multicolumn{2}{c}{\textbf{Head}} \\
    \cmidrule(lr){2-3} \cmidrule(lr){4-5} \cmidrule(lr){6-7}
     & N@10 & H@10 & N@10 & H@10 & N@10 & H@10 \\
    \midrule
    \textbf{\name} & \textbf{0.4272} & \textbf{0.6692} & \textbf{0.1306} & \textbf{0.2693} & \textbf{0.5483} & \textbf{0.8324} \\
    \textit{w/o} FN & 0.4123 & 0.6605 & 0.1013 & 0.2347 & 0.5404 & 0.8208 \\
    \textit{w/o} $\mathcal{L}_{CA}$ & 0.4044 & 0.6455  & 0.1067 & 0.2161 & 0.4920  & 0.7719 \\
    \textit{w/o} MCA & 0.4072 & 0.6523 & 0.1105 & 0.2190 & 0.5218 & 0.8077 \\
    \textit{w/o} $\mathcal{L}_{MSG}$ & 0.4202 & 0.6667  & 0.1091 & 0.2285 & 0.5398  & 0.8258  \\
    \bottomrule
  \end{tabular}
  }
\end{table}

\subsection{Ablation Study}
To assess the contribution of each proposed component, we conduct an ablation study by selectively removing individual modules while keeping the remaining architecture unchanged. The variants are defined as follows:
(1) \textit{w/o Fusion Network}: Removes the proposed multi-granularity fusion network and replaces it with fixed weights across different granularities.
(2) \textit{w/o} $\mathcal{L}_{CA}$: Eliminates the code-guided alignment loss.
(3) \textit{w/o Multi-granularity Cross Attention}: Removes the multi-granularity cross-attention module to examine the importance of information transfer between the HID and SID branches.
(4) \textit{w/o} $\mathcal{L}_{MSG}$: Removes the masked sequence granularity loss to evaluate its contribution. Based on the empirical observations from \textbf{Table}~\ref{tab:ablation}, we summarize the findings below:
\begin{itemize}[leftmargin=*]

\item \textbf{Effectiveness of Fusion and MSG Loss.}  
Both the fusion network and $\mathcal{L}_{MSG}$ improve the semantic representation quality. The improved SID embeddings lead to better performance on Tail items, demonstrating the importance of learning adaptive and robust multi-granular semantics.

\item \textbf{Role of Cross-branch Interaction.}  
Removing the MCA module results in a notable performance degradation on Head items. This confirms that cross-branch information transfer is essential, as it enables the HID branch to selectively absorb beneficial multi-granular semantics without losing identifier uniqueness.

\item \textbf{Importance of Code-guided Alignment.}  
The removal of $\mathcal{L}_{CA}$ results in declines across both groups, indicating that our alignment strategy effectively facilitates accurate, noise-resistant collaborative information sharing among semantically related items.
\end{itemize}

These results demonstrate that \name~framework successfully leverages the complementary strengths of HID and SID, thereby addressing the aforementioned problems outlined in \textbf{Section}~\ref{sec:intro}.

\subsection{Hyperparameter Analysis}
\begin{figure}[!t]
	\centering
	\includegraphics[width = \linewidth]{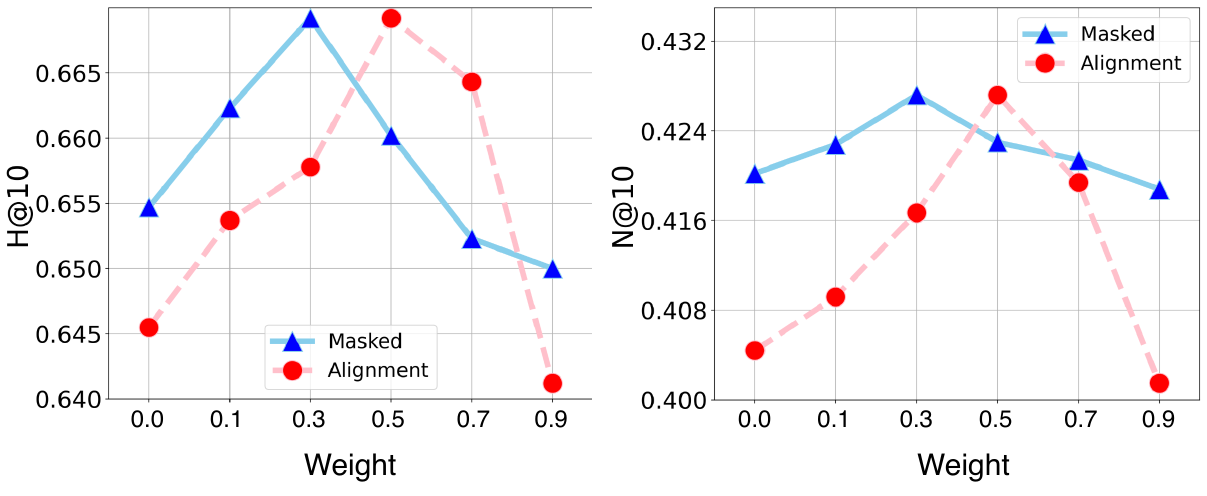}
	\caption{Hyper-parameter Results on Yelp Dataset. ``Masked'' represents the $\mathcal{L}_{MSG}$ while ``Alignment'' represents the $\mathcal{L}_{CA}$.}
        \label{fig:hyper}
\end{figure}
We analyze the sensitivity of two key hyperparameters: the weight of the alignment loss $\beta$ and the weight of the granularity-masked loss $\gamma$. The results on the Yelp dataset are presented in \textbf{Figure}~\ref{fig:hyper}. Both hyperparameters exhibit clear and consistent trends.

\subsubsection{\textbf{Alignment Weight $\beta$}}
Performance increases as $\beta$ grows and reaches the optimum around $\beta = 0.5$.
A small value of $\beta$ compromises the code-guided alignment, causing an insufficient transfer of collaborative signals to semantically similar items. Conversely, an excessively large $\beta$ causes the semantic space to over-align to the collaborative space, introducing noise and harming generalization.

\subsubsection{\textbf{Granularity Weight $\gamma$}}
From the \textbf{Figure}~\ref{fig:hyper}, the performance for our masked sequence granularity loss peaks at $\gamma = 0.3$. As an auxiliary regularizer designed to enforce internal consistency, $\mathcal{L}_{MSG}$ requires a balanced weight: values that are too low fail to ensure representation robustness, while excessive weights may interfere with the primary recommendation task.

\subsection{In-depth Analysis for $\mathcal{L}_{CA}$}
To further validate the effectiveness of our code-guided strategy in mitigating the noisy collaborative sharing problem, we analyze the influence of two key design factors in $\mathcal{L}_{CA}$: the code-matching threshold $p$ and the context window size $o$.

\subsubsection{\textbf{Impact of Code Matching Threshold}}
\textbf{Table}~\ref{tab:abl_p} shows that using $p = 1$, which considers only the coarsest semantic layer, leads to notable performance degradation. This observation indicates that overly coarse categories tend to group many weakly related items together, thereby introducing noise into the positive set. Increasing the threshold to $p = 3$ yields significant improvements across all metrics, especially for Tail items. This demonstrates that deeper semantic matching is crucial for filtering out noise and ensuring that only truly similar items share collaborative signals.

\subsubsection{\textbf{Impact of Context Window Size}}
As shown in \textbf{Table}~\ref{tab:abl_o}, expanding the co-occurrence window improves performance up to $o = 3$, confirming the utility of incorporating local, sequentially related items as additional positives. However, when the window is enlarged to $o = 5$, the model begins to include less relevant items, which results in a performance drop.

These results verify that our design, which combines multi-level semantic matching with local collaborative context, is essential for accurate and noise-robust alignment within $\mathcal{L}_{CA}$.

\begin{table}[t]
  \centering
  \caption{Impact of the semantic code-matching threshold $p$ on the Yelp dataset. Best results are highlighted in bold.}
  \label{tab:abl_p}
  \resizebox{\linewidth}{!}{
  \begin{tabular}{lcccccc}
    \toprule
    \multirow{2}{*}{\textbf{code num}} & \multicolumn{2}{c}{\textbf{Overall}} & \multicolumn{2}{c}{\textbf{Tail}} & \multicolumn{2}{c}{\textbf{Head}} \\
    \cmidrule(lr){2-3} \cmidrule(lr){4-5} \cmidrule(lr){6-7}
     & N@10 & H@10 & N@10 & H@10 & N@10 & H@10 \\
    \midrule
   Removed & 0.4095 & 0.6505  & 0.1177 & 0.2211 & 0.4970  & 0.7769 \\
    1 &0.4059  &0.6473  &0.1058  &0.2044  &0.4955  &0.7712  \\
    2 &0.4199  &0.6582  &0.1204  &0.2591  &0.5385  &0.8247  \\
    3 (Ours) & \textbf{0.4272} & \textbf{0.6692} & \textbf{0.1306} & \textbf{0.2693} & \textbf{0.5483} & \textbf{0.8324} \\

    \bottomrule
  \end{tabular}
  }
\end{table}
\begin{table}[t]
  \centering
  \caption{Impact of the context window size $o$ on the Yelp dataset. Best results are highlighted in bold.}
  \label{tab:abl_o}
  \resizebox{\linewidth}{!}{
  \begin{tabular}{lcccccc}
    \toprule
    \multirow{2}{*}{\textbf{context}} & \multicolumn{2}{c}{\textbf{Overall}} & \multicolumn{2}{c}{\textbf{Tail}} & \multicolumn{2}{c}{\textbf{Head}} \\
    \cmidrule(lr){2-3} \cmidrule(lr){4-5} \cmidrule(lr){6-7}
     & N@10 & H@10 & N@10 & H@10 & N@10 & H@10 \\
    \midrule
   Removed & 0.4148 & 0.6522  & 0.1176 & 0.2230 & 0.5072  & 0.7911 \\
    1 &0.4190  &0.6523  &0.1208  &0.2294  &0.5205  &0.8062  \\
    3 (Ours) & \textbf{0.4272} & \textbf{0.6692} & \textbf{0.1306} & \textbf{0.2693} & \textbf{0.5483} & \textbf{0.8324} \\
    5 &0.4205  &0.6637  &0.1244  &0.2638  &0.5441  &0.8299  \\

    \bottomrule
  \end{tabular}
  }
\end{table}
\begin{figure}[!t]
	\centering
	\includegraphics[width = \linewidth]{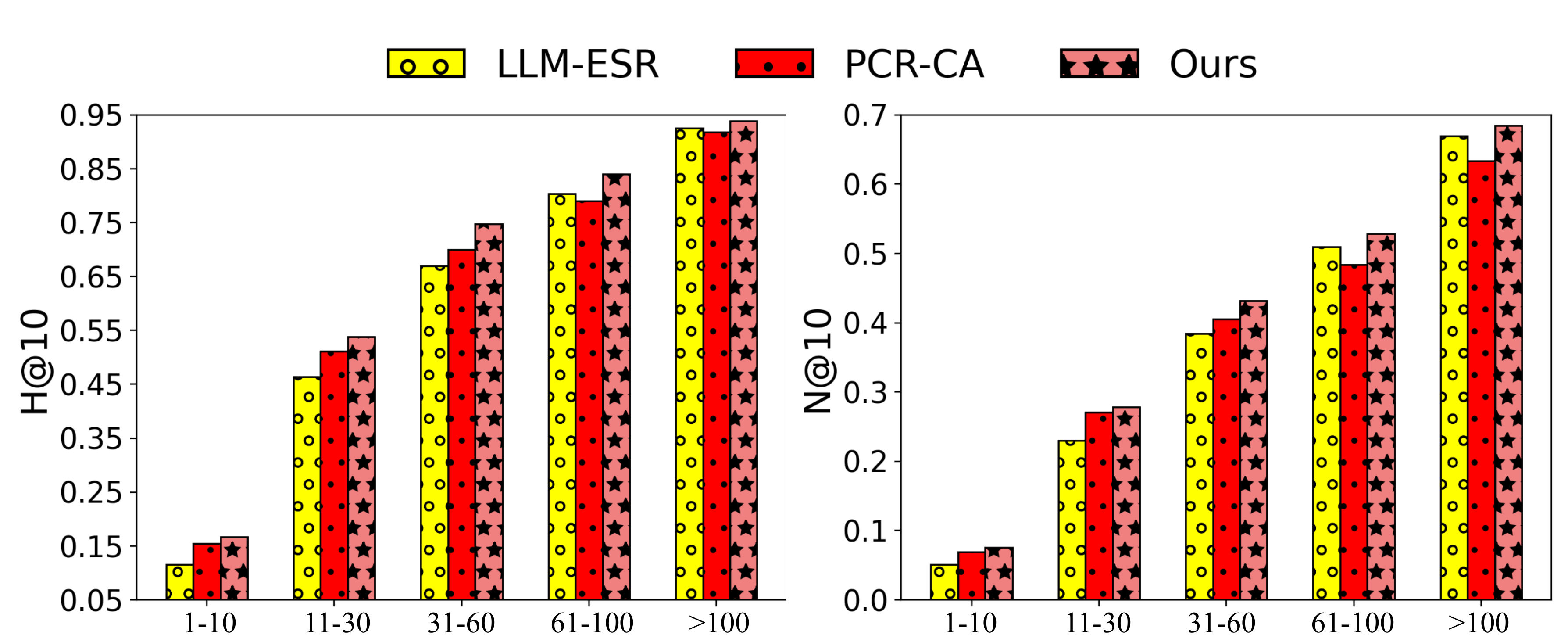}
	\caption{Detailed results in different item groups on the Yelp.}
        \label{fig:group}
\end{figure}
\subsection{Group Analysis}
For a more fine-grained analysis, we further divide the original head–tail item group in \textbf{Table}~\ref{tab:overall_performance} into five groups according to item popularity. The results are illustrated in \textbf{Figure}~\ref{fig:group}. From the figure, we observe that no matter for $H@10$ or $N@10$, PCR-CA, which adopts a hybrid embedding mechanism, achieves a clear performance gain over the LLM-enhanced method (\ie, LLM-ESR) in groups with popularity lower than $60$ (the first three grouped columns). In addition, its performance in groups with popularity greater than $60$ (the last two grouped columns) is almost on par with that of LLM-ESR, validating the benefit of hybrid embedding.

In contrast, our method further leverages a dual-branch architecture together with a dual-level alignment strategy to achieve additional optimization. As a result, it consistently outperforms all existing methods across all popularity groups. 
\subsection{Generality Validation}
\subsubsection{\textbf{Quantization Mechanism Analysis}}
In this section, we compare the performance of \name~using SID generated by three commonly adopted quantization mechanisms, \ie VQ~\cite{vq}, PQ~\cite{pq}, RQ~\cite{RQVAE}, and a novel tokenizer called OPMQ from another generative recommendation framework~\cite{STORE}. For the detailed results and analysis, please refer to \textbf{Appendix}~\ref{sec:quanti_general}.
\subsubsection{\textbf{Backbone Analysis}}
To further verify the model-agnostic nature of our framework, we evaluate \name~under two widely adopted SRS backbones, \ie GRU4Rec and BERT4Rec, against two competitive hybrid baselines, \ie PCR-CA and LLM-ESR. Please refer to \textbf{Appendix}~\ref{sec:backbone_general} for details.

\subsection{Semantic Code Analysis}
In this section, we present how our proposed \name~performs with SID in different qualities. We conduct our experiments on two aspects, \ie code layers and codebook size. Since there are too many combinations to choose from, we only test the SID with representative settings. The results are reported in Table \ref{tab:codeset}.
\begin{table}[t]
  \centering
  \caption{Performance comparison of different SID settings on the Yelp Dataset. "Coll." represents the collision rate of SID, while "Util." represents the utilization rate.}
  \label{tab:codeset}
  \resizebox{\linewidth}{!}{
  \begin{tabular}{lcccccccc}
    \toprule
    \multirow{2}{*}{\textbf{Settings}} & \multicolumn{2}{c}{\textbf{Overall}} & \multicolumn{2}{c}{\textbf{Tail}} & \multicolumn{2}{c}{\textbf{Head}} & \multirow{2}{*}{\textbf{Coll.}\%} & \multirow{2}{*}{\textbf{Util.}\%} \\
    \cmidrule(lr){2-3} \cmidrule(lr){4-5} \cmidrule(lr){6-7}
      & N@10 & H@10 & N@10 & H@10 & N@10 & H@10 & & \\
    \midrule
    3x128 & 0.4145 & 0.6510 & 0.1152 & 0.2480 & 0.5365 & 0.8150 & 33.92 & 0.54 \\
    3x256 & 0.4192 & 0.6585 & 0.1215 & 0.2565 & 0.5402 & 0.8215 & 29.92 & 0.068 \\
    3x512 & 0.4235 & 0.6640 & 0.1268 & 0.2630 & 0.5445 & 0.8270 & 25.85 & 0.008 \\
    4x128 & 0.4272 & 0.6692 & 0.1306 & 0.2693 & 0.5483 & 0.8324 & 22.28 & 0.004 \\
    4x256 & \textbf{0.4338} & \textbf{0.6785} & \textbf{0.1385} & \textbf{0.2790} & \textbf{0.5540} & \textbf{0.8410} & 11.87 & <0.001 \\
    \bottomrule
  \end{tabular}
  }
\end{table}
From \textbf{Table}~\ref{tab:codeset}, we observe that expanding codebook capacity yields consistent improvements, which stems from the significant reduction in collision rates. Lower collision rates mitigate the semantic ambiguity, thereby enhancing the distinctiveness of item representations. However, this performance gain comes at the cost of codebook redundancy. As observed, the $4\times256$ setting suffers from an extremely low utilization rate compared to smaller settings. Therefore, to strike a balance between minimizing collision rates (to ensure distinctiveness) and maintaining a reasonable utilization rate (to ensure efficiency), we set $4\times128$ as the default setting.

\subsection{Online Experiment}
We conducted a large-scale online A/B test on the dual-column feed of a leading content platform serving over millions of daily active users. The proposed method, \name, was deployed as a candidate generation strategy within the recall stage of the recommendation pipeline. Under strict traffic isolation protocols, 14\% of live user traffic was equally allocated to treatment (integrating \name) and control (production baseline) groups (7\% per group). Evaluation centered on two business-critical metrics: \textbf{Advertising Value Equivalency (ADVV)}, the product of conversions and bid price, quantifying delivered value to advertisers;  
\textbf{Advertising Cost (COST)}, total advertiser spend (i.e., platform ad revenue), reflecting monetization scale and resource allocation efficiency. Together, these metrics holistically assess the health of the advertising ecosystem and operational effectiveness.
Results demonstrate statistically significant relative gains ($p < 0.001$) with \name: \textbf{+0.89\% in ADVV} and \textbf{+0.59\% in COST}, with no degradation in core user engagement metrics (e.g., dwell time, click-through rate). 
This dual improvement confirms that \name~ enhances advertiser benefits and platform monetization efficiency without compromising user experience.
Our findings deliver reproducible, actionable insights for industrial-scale recommendation systems, validating \name’s effectiveness in harmonizing HID and SID representations.

\section{Related Work}

\subsection{Sequential Recommender Systems} 
The primary goal of Sequential Recommender Systems (SRS) is to model dynamic user preferences based on historical interaction sequences for next-item prediction~\cite{stream,e-commerce,zhang2025glint,wang2025star,liu2025sigma,zhao2018deep,zhao2018recommendations,liu2023exploration,representation1,knowledge_transfer}. Early methodologies adapted neural architectures to this task, including RNN-based GRU4Rec~\cite{GRU4Rec}, CNN-based Caser~\cite{Caser}, and MLP-based models~\cite{li2022mlp4rec,li2023automlp,liang2023mmmlp}. Following the breakthrough of Transformers in natural language processing, self-attention mechanisms~\cite{attention}, exemplified by SASRec~\cite{SASRec} and BERT4Rec~\cite{Bert4rec}, have become the dominant paradigm due to their superior ability to capture long-range dependencies. Despite their success, conventional ID-based frameworks face intrinsic limitations. First, for tail items with sparse interactions, the model often struggles to learn reliable representations~\cite{sparsity}. While prior studies attempt to leverage co-occurrence patterns to enhance tail items~\cite{Melt,CITIES}, these signals can be unreliable in real-world scenarios, leading to the \textbf{Noisy Collaborative Sharing} problem where tail items inherit misleading signals from unrelated popular items~\cite{longtail_session}. 

To mitigate data sparsity, recent research has incorporated Large Language Models (LLM) to encode textual attributes into item representations~\cite{LLM-ESR,surveyllmesr,LLMemb,liu2025llm,gao2025llm4rerank,zhang2025notellm,li2022gromov}. However, most existing approaches compress detailed textual descriptions into a single dense vector. As argued in Section~\ref{sec:intro}, this ``flat'' representation suffers from a single-granularity bottleneck, entangling coarse-grained semantics with fine-grained nuances. Consequently, the model fails to distinguish subtle discrepancies between semantically similar items, leading to the \textbf{Semantic Homogeneity} problem. Our proposed \name~addresses these issues by introducing harmonized Semantic and Hash IDs, leveraging a dual-level alignment strategy to refine both representation spaces.

\subsection{Semantic IDs Enhanced Sequential Recommendation} 
To overcome the limitations of Hash IDs (HID) and dense embeddings, researchers have investigated the paradigm of Semantic IDs (SID). SID are typically generated by decomposing dense semantic features (\eg from LLMs) into discrete code sequences via vector quantization techniques like RQ-VAE~\cite{RQVAE}. Theoretically, this discretization process naturally unfolds the flat embedding into multi-granular semantic views, offering a structural advantage for recommendation. Methods such as VQRec~\cite{VQRec}, SPM-SID~\cite{SPM_SID}, PG-SID~\cite{prefixgram_SID}, and HiD-VAE~\cite{HiD-VAE} leverage and enhance the code-sharing nature of SID, where semantically similar items share the same code. By aggregating signals based on semantic content rather than unreliable co-occurrence, these methods aim to mitigate \textbf{Noisy Collaborative Sharing}. Moreover, approaches such as CCFRec~\cite{CCFRec} and PSRQ~\cite{PSRQ} leverage the hierarchical structure of SID to capture item semantics to alleviate the \textbf{Semantic Homogeneity}.

However, the effective utilization of SID remains a challenge. Current state-of-the-art methods typically adopt a substitution strategy~\cite{ADA-SID,VQRec} or fuse identifiers via simple concatenation or contrastive learning~\cite{PCRCA,MME-SID,Meta_concat,SMILE}. These straightforward integration strategies overlook a critical trade-off: the \textbf{Collaborative Overwhelming Phenomenon}. The quantization process inevitably introduces code collisions and lossy compression, which compromises the uniqueness of item identifiers. In pure or lightly fused SID frameworks, shared semantic codes tend to "overwhelm" the unique distinctiveness required for popular (head) items, resulting in performance degradation where ID uniqueness is crucial. In contrast, \name~employs a dual-branch framework that harmonizes semantic and hash IDs. By synergizing multi-granularity cross-attention with a Dual-level Alignment strategy, our approach effectively balances the uniqueness of HID with the semantic generalization of SID.
\section{Conclusion}
In this work, we investigated the fundamental trade-off between identifier uniqueness and semantic generalization in sequential recommendation, and formally identified the \textbf{Collaborative Overwhelming} phenomenon. To address the performance imbalance between head and tail items, we proposed \textbf{\name}, a dual-branch framework that harmonizes Hash IDs (HID) and Semantic IDs (SID). By integrating a multi-granularity fusion network with a multi-granularity cross attention Network, \name~effectively captures fine-grained semantic distinctions while preserving the collaborative specificity necessary for popular items, alleviating the \textbf{Semantic Homogeneity}. Moreover, our dual-level alignment strategy bridges the semantic and collaborative spaces, enabling long-tail items to borrow high-quality signals and mitigating the \textbf{Noisy Collaborative Sharing}. Extensive experiments online and offline validate the superiority of our proposed \name. Importantly, it breaks the performance bottleneck by achieving substantial gains on tail items without sacrificing performance on head items.
\begin{acks}
This research was partially supported by National Natural Science Foundation of China (No.62502404), Hong Kong Research Grants Council (Research Impact Fund No.R1015-23, Collaborative Research Fund No.C1043-24GF, General Research Fund No. 11218325), Institute of Digital Medicine of City University of Hong Kong (No.9229503), Huawei (Huawei Innovation Research Program), Tencent (Tencent Rhino-Bird Focused Research Program, Tencent University Cooperation Project), Kuaishou (CCF-Kuaishou Large Model Explorer Fund No. 2025008, Kuaishou University Cooperation Project), Didi (CCF-Didi Gaia Scholars Research Fund), and Bytedance.
\end{acks}
\bibliographystyle{ACM-Reference-Format}
\balance
\bibliography{8Reference}
\clearpage
\appendix
\section{Vector Quantization} \label{sec:quan}
In this section, we will present the detailed procedure of the quantization mechanism (RQ-VAE) we adopt.
\subsection{Item Embedding Generation}
For the RQ-VAE input, we first design a prompt to summarize the item's semantic features. Specifically, we construct the template by combining items' attributes, \ie, Title, Brand, Date, Price, Categories, and Description as follows:
\begin{tcolorbox}[
        colframe=gray!50!,
        width=1\linewidth,
        arc=1mm, 
        auto outer arc,
        title={Item Prompt Template},
        breakable,]
        
        The \ul{<dataset>} item has the following attributes: \\
        name is \ul{<TITLE>}; brand is \ul{<BRAND>}; date is \ul{<DATE>}; price is \ul{<PRICE>}. \\
        The item has the following features: \ul{<FEATURE>}. \\
        The item has the following descriptions: \ul{<DESCRIPTION>}. 
        
\end{tcolorbox}
Then, the LLM embeddings are derived using the "text-ada-embedding-002" API to process the template for each item.
\subsection{RQ-VAE}
Aiming to tokenize and generate semantic IDs in a hierarchical manner, we adopt the commonly used quantization mechanism, \ie Residual Quantized Variational Autoencoder (RQ-VAE)~\cite{RQVAE} as our SID generator. Specifically, for semantic embeddings $\boldsymbol{e}_{LLM}$ derived from a language model, we leverage an existing RQ-VAE framework~\cite {GRID} to quantize them into codes using \textit{L}-level codebooks. 
\begin{equation}
\begin{aligned}
    \boldsymbol{z} &= \operatorname{Encoder}(\boldsymbol{e}_{LLM}) \\
    c^l &= \arg \min_j ||\boldsymbol{r}_{l-1} - \boldsymbol{e}^j_{c}|| \\
    \boldsymbol{r}_l &= \boldsymbol{r}_{l-1} - \boldsymbol{e}_{c}^{l} 
\end{aligned}
\end{equation}
where $c^l$ represents the assigned
code for $l$-th code layer, $\boldsymbol{r}_{l-1}$ is the residual from the last layer, $\boldsymbol{r}_{0} = \boldsymbol{z}$, and $||\cdot||$ is \textit{L2} norm. The quantized embedding $\boldsymbol{\hat{z}} = \sum^{L}_{l=1} \boldsymbol{e}_c^{l}$ for any item.
Then, to construct the reconstruction loss, we need to expand the dimension back to the original embedding $\boldsymbol{e}_{LLM}$ using a decoder. The overall loss function can be formulated as:
\begin{equation}
    \begin{gathered}
        \mathcal{L} = \mathcal{L}_{Recon} + \mathcal{L}_{RQ-VAE} \\
        \boldsymbol{\hat{e}_D} = \operatorname{Decoder}(\boldsymbol{\hat{z}}) \\
        \mathcal{L}_{Recon} = \|\boldsymbol{e}_{LLM} -\boldsymbol{\hat{e}_D} \| \\ 
        \mathcal{L}_{RQ-VAE} = \sum_{l=1}^{L} \|\operatorname{SG}(\boldsymbol{r}_{l-1}-\boldsymbol{e}_c^l)\|^2 + \alpha\|\boldsymbol{r}_{l-1} - \operatorname{SG}(\boldsymbol{e}_c^l)\|^2     
    \end{gathered}
\end{equation}
Where $\operatorname{SG}$ denotes the stop gradient operation, and $\alpha$ is a hyper-parameter used for balancing the RQ-VAE loss and reconstruction loss. Following previous work, we set $\alpha$ to $1.0$.
\section{Descriptions and Settings}

\subsection{Data Descriptions}
\label{sec:dataset}
In this section, we present the detailed statistics of the selected public datasets, \ie Yelp, Amazon Beauty, and Amazon Instrument. For data preprocessing, we follow the previous SRS works~\cite{LLM-ESR,liu2025sigma}. The statistics after preprocessing are presented in Table~\ref{tab:dataset}.
\begin{table}[t]
\centering
\caption{The statistics of datasets}
\label{tab:dataset}
\begin{tabular}{ccccc}
\toprule
Dataset & \# Users & \# Items & Sparsity & Avg.length \\ 
\midrule
Yelp & 15,720 & 11,383 & 99.89\% & 12.23 \\
Beauty & 52,204 & 57,289 & 99.92\% & 7.56 \\
Instrument & 40,644 & 30,676 & 99.97\% & 8.01\\
\bottomrule
\end{tabular}
\end{table}
\begin{algorithm}[t]
    \caption{Training and Inference Procedures of \name.}
    \label{alg:hiser_alg}
    \begin{algorithmic}[1]
        \Require User set $\mathcal{U}$, Item set $\mathcal{I}$
        \State Define the backbone SRS encoders $f_{\theta^{sid}}$ and $f_{\theta^{hid}}$.
        \State Define the hyperparameters for weights of $\mathcal{L}_{CA}$ and $\mathcal{L}_{MSG}$.
        \State Pre-compute semantic embedding $\boldsymbol{E}_{LLM}$ via LLM and codebook $\boldsymbol{E}_C$ via RQ-VAE.
    \Statex \textbf{Training}
        \State Initialize HID item embeddings using dimension-reduced $\boldsymbol{E}_{LLM}$~\cite{PCA}.
        \State Initialize semantic code embeddings by looking up entries in $\boldsymbol{E}_{C}$.
        \For{each batch of users $\mathcal{U}_B \subset \mathcal{U}$}
            \State Retrieve embedding sequences for different code granularities.
            \State Compute granularity weights $\alpha_l$ via Equation~\eqref{equ:score}.
            \State Derive the final SID item embedding sequence $\mathcal{E}^{sid}$ via Equation~\eqref{equ:weights}.
            \State Derive the fused HID embedding sequence $\mathcal{E}^{f}$ via Equation~\eqref{equ:MCA}.
            \State Generate user representations: $\boldsymbol{u}^{sid}$ (via $f_{\theta^{sid}}$) and $\boldsymbol{u}^{hid}$ (via $f_{\theta^{hid}}$).
            \State Compute prediction scores for ground-truth and negative items via Equation~\eqref{equ:rec}.
            \State Compute the main ranking loss $\mathcal{L}_{rec}$ via Equation~\eqref{equ:recloss}.
            \State Compute alignment losses: $\mathcal{L}_{CA}$ (Eq.~\eqref{equ:lca}) and $\mathcal{L}_{MSG}$ (Eq.~\eqref{equ:lmsg}).
            \State Update parameters by minimizing $\mathcal{L}_{total} = \mathcal{L}_{rec} + \beta\mathcal{L}_{CA} + \gamma \mathcal{L}_{MSG}$.
        \EndFor
    \Statex \textbf{Inference}
        \State Load trained parameters, including $\boldsymbol{E}_{LLM}$, $\boldsymbol{E}_C$, and projection weights.
        \For{each batch of users $\mathcal{U}_B \subset \mathcal{U}$}
            \State Generate user representations $\boldsymbol{u}^{sid}$ and $\boldsymbol{u}^{hid}$ following the training logic.
            \State Compute prediction scores via Equation~\eqref{equ:rec} to generate top-$K$ recommendations.
        \EndFor
    \end{algorithmic}
\end{algorithm}
\subsection{Implementation Details} \label{sec:imple}
In this section, we will present the implementation details of our \name. For the quantization stage, since our framework is quantization-agnostic, we use the common RQ-VAE framework with a codebook hidden state size of 128 and 3 code layers. Other details follow pioneering work~\cite{GRID}. For the recommendation stage, we set a batch size of 128 and a maximum user sequence of 200. Other details follow the LLM-ESR framework~\cite{LLM-ESR}.

\subsection{Training \& Inference Algorithm}
\label{sec:train_infer}
To provide a comprehensive overview of the data flow within our \name~framework, we detail the complete training and inference procedures in Algorithm~\ref{alg:hiser_alg}. 

The process is initiated by defining the backbone and necessary hyperparameters (lines 1-2). Crucially, to bridge the gap between collaborative signals and semantic knowledge, we pre-compute the semantic embeddings using the LLM and generate a discrete codebook via the RQ-VAE module (line 3). During the training phase, after initializing the HID item embeddings and semantic code entries (lines 4-5), the model iterates through user batches (line 6) to retrieve embedding sequences across different code granularities (line 7). Subsequently, granularity weights are dynamically computed (line 8) to derive the final SID item embedding sequence and the fused HID embedding sequence (lines 9-10). With these sequences prepared, the backbone encoders generate specific user representations (line 11). The optimization process involves calculating prediction scores (line 12), constructing the objective by combining the main ranking loss with alignment losses (lines 13-14), and then performing parameter updating (line 15). As for the inference, the trained parameters are loaded (line 17) to generate user representations and compute prediction scores, ultimately yielding the top-$K$ recommendation list (lines 18-20).
\begin{table*}[t] 
\centering
\caption{Performance comparison on the Yelp dataset with different backbones. The best results are highlighted in bold.}
\label{tab:model}
\begin{tabular}{llcccccc}
\toprule
\multirow{2}{*}{\textbf{Backbone}} & \multirow{2}{*}{\textbf{Model}} & \multicolumn{2}{c}{\textbf{Overall}} & \multicolumn{2}{c}{\textbf{Tail Items}} & \multicolumn{2}{c}{\textbf{Popular Items}} \\
\cmidrule(lr){3-4} \cmidrule(lr){5-6} \cmidrule(lr){7-8}
 & & N@10 & H@10 & N@10 & H@10 & N@10 & H@10 \\
\midrule
\multirow{3}{*}{GRU4Rec} 
 & PCR-CA & 0.3613 & 0.5988 & 0.0745 & 0.1889 & 0.4311 & 0.7204 \\
 & LLM-ESR & 0.3627 & 0.6075 & 0.0482 & 0.0952 & 0.4491 & 0.7338 \\
 & \textbf{Ours} & \textbf{0.3804} & \textbf{0.6239} & \textbf{0.0983} & \textbf{0.2058} & \textbf{0.4634} & \textbf{0.7470} \\
\midrule
\multirow{3}{*}{BERT4Rec} 
 & PCR-CA & 0.4173 & 0.6604 & 0.0720 & 0.1689 & 0.5346 & 0.8175 \\
 & LLM-ESR & 0.4205 & 0.6635 & 0.0503 & 0.1247 & 0.5444 & 0.8223 \\
 & \textbf{Ours} & \textbf{0.4298} & \textbf{0.6724} & \textbf{0.0991} & \textbf{0.1735} & \textbf{0.5545} & \textbf{0.8344} \\
\bottomrule
\end{tabular}
\end{table*}
\section{Supplementary Experiments}
\subsection{Backbone Generality Analysis} \label{sec:backbone_general}
To validate the generality of our \name, we also test the performance under different SRS backbones, \ie GRU4Rec~\cite{GRU4Rec} and BERT4Rec~\cite{Bert4rec}. 
The results shown in Table~\ref{tab:model} demonstrate that, across both RNN-based and Transformer-based architectures, \name~consistently surpasses all baselines. This confirms that our framework is not restricted to a specific encoder design and can function as a universal, plug-and-play enhancement module for various SRS.
\begin{table}[t]
  \centering
  \caption{Performance comparison of \name~with different quantization mechanisms on the Yelp Dataset. Best results are highlighted in bold, second are underlined.}
  \label{tab:quan_results}
  \begin{tabular}{lcccccc}
    \toprule
    \multirow{2}{*}{\textbf{Model}} & \multicolumn{2}{c}{\textbf{Overall}} & \multicolumn{2}{c}{\textbf{Tail}} & \multicolumn{2}{c}{\textbf{Head}} \\
    \cmidrule(lr){2-3} \cmidrule(lr){4-5} \cmidrule(lr){6-7}
     & N@10 & H@10 & N@10 & H@10 & N@10 & H@10 \\
    \midrule
    VQ + & 0.4091 & 0.6531 & 0.0943 & 0.1986 & 0.5018 & 0.7869 \\
    PQ + & 0.4124 & 0.6597 & 0.1091 & 0.2285 & 0.5118 & 0.7958 \\
    RQ + & \uline{0.4272} & \uline{0.6692} & \uline{0.1306} & \uline{0.2693} & \uline{0.5483} & \uline{0.8324} \\
    OPMQ + & \textbf{0.4334} & \textbf{0.6801} & \textbf{0.1372} & \textbf{0.2751} & \textbf{0.5510} & \textbf{0.8409} \\
    \bottomrule
  \end{tabular}
\end{table}

\subsection{Quantization Mechanism Generality Analysis} \label{sec:quanti_general}
In this section, we will analyze the performance of our \name~ in detail. As shown in Table~\ref{tab:quan_results}, \name~maintains competitive performance under all configurations, demonstrating the robustness and general applicability of our design.

We further provide insights into the performance differences:
\begin{itemize}[leftmargin=*] 
    \item \textbf{OPMQ} achieves the strongest performance because of the orthogonalized multi-experts structure, which enforces different experts to learn the corresponding code embedding for different features, leading to a more fine-grained representation compared to normal quantization mechanisms.
    \item \textbf{RQ and PQ} provide moderate improvements because they process the embedding space by leveraging the residual or parallel structure to acquire different views of the item's features.
    \item \textbf{VQ} performs the weakest because mapping each item to a single discrete code causes severe semantic collapse, reducing item uniqueness across both head and tail groups.
\end{itemize}
Despite these differences, all variants achieve consistently strong results, even under the least expressive VQ setting. This demonstrates that \name~can effectively exploit semantic signals regardless of the underlying quantization mechanism, confirming the generalizability and plug-and-play nature of our framework.

\end{document}